\documentclass[twocolumn,runningheads]{svjour2}
\smartqed  
\usepackage{graphicx}
\journalname{Astrophysics and Space Science}
\begin{document}

\title{Anomalous X-ray pulsars: Persistent States with Fallback Disks
\thanks{We acknowledge support from the Astrophysics and Space
Forum at Sabanc{\i} University. 
S.\c{C}. acknowledges support from the FP6 Marie Curie Reintegration Grant, INDAM.}}



\author{\"{U}. Ertan, M. A. Alpar, M. H. Erkut, K. Y. Ek\c{s}i,\\  
\& \c{S}. \c{C}al{\i}\c{s}kan }


\institute{\"{U}. Ertan \at
        Sabanc\i\ University, 34956, Orhanl\i\, Tuzla, \.Istanbul, Turkey \\                          
              \email{unal@sabanciuniv.edu}           
           \and
           M. A. Alpar \at
        Sabanc\i\ University, 34956, Orhanl\i\, Tuzla, \.Istanbul, Turkey \\                  \email{alpar@sabanciuniv.edu}   
	           \and
           M. H. Erkut \at
        Sabanc\i\ University, 34956, Orhanl\i\, Tuzla, \.Istanbul, Turkey \\                   \email{hakane@sabanciuniv.edu}      
	           \and
           K. Y. Ek\c{s}i  \at
	\.Istanbul Technical University (\.{I}T\"{U}), \.Istanbul, Turkey  
               \email{eksi@itu.edu.tr}       }

\date{Received: date / Accepted: date}

\maketitle

\begin{abstract}

The anomalous X-ray pulsar 4U 0142+61 was recently detected in the mid  infrared bands with the SPITZER Observatory (Wang, Chakrabarty \& Kaplan 2006). This observation is the first instance for a disk around an AXP. From a reanalysis of optical and infrared data, we show that the observations indicate that the disk is likely to be an active disk rather than a passive dust disk beyond the light cylinder, as proposed in the discovering paper. Furthermore, we show that the irradiated accretion disk model can also account for all the optical and infrared observations of the anomalous X-ray pulsars in the persistent state. 


\end{abstract}

\def\la{\raise.5ex\hbox{$<$}\kern-.8em\lower 1mm\hbox{$\sim$}}
\def\ga{\raise.5ex\hbox{$>$}\kern-.8em\lower 1mm\hbox{$\sim$}}
\def\be{\begin{equation}}
\def\ee{\end{equation}}  
\def\ba{\begin{eqnarray}}
\def\ea{\end{eqnarray}}  
\def\be{\begin{equation}}
\def\ee{\end{equation}}  
\def\ba{\begin{eqnarray}}
\def\ea{\end{eqnarray}}  
\def\m{\mathrm}
\def\d{\partial}
\def\R{\right}
\def\L{\left}
\def\a{\alpha}
\def\Mdot*{\dot{M}_*}
\def\Mdotin{\dot{M}_{\mathrm{in}}}
\def\Mdot{\dot{M}}
\def\Lin{L_{\mathrm{in}}}
\def\Rin{R_{\mathrm{in}}}
\def\Rout{R_{\mathrm{out}}}
\def\Rout{R_{\mathrm{out}}}
\def\Ldisk{L_{\mathrm{disk}}}
\def\Lx{L_{\mathrm{x}}}
\def\dEb{\delta E_{\mathrm{burst}}}
\def\dEx{\delta E_{\mathrm{x}}}
\def\Bb{\beta_{\mathrm{b}}}
\def\Be{\beta_{\mathrm{e}}}
\def\Rc{\R_{\mathrm{c}}}
\def\dMin{\delta M_{\mathrm{in}}}
\def\dM*{\delta M_*}
\def\Teff{T_{\mathrm{eff}}}
\def\Tirr{T_{\mathrm{irr}}}
\def\Firr{F_{\mathrm{irr}}}
\def\Av{A_{\mathrm{V}}}
\def\p{\propto}
\def\m{\mathrm}
\def\Ks{K_{\mathrm{s}}}

\section{Introduction}
\label{intro}

Anomalous X-ray Pulsars (AXPs) and Soft Gamma-ray Repeaters (SGRs) (Mereghetti et al. 2002; Hurley 2000; 
Woods $\&$ Thompson 2006) are characterized mainly by their X-ray luminosities 
($\Lx \sim 10^{34}-10^{36}$ erg s$^{-1}$), which are orders of magnitude higher than their rotational powers $\dot{E}_{rot}= I \Omega \dot{\Omega}$. Their spin periods are clustered to a very narrow range ($5 - 12$ s). Most of  SGRs and AXPs show repetitive, short (\la 1s) super-Eddington bursts with luminosities up to 
$10^{42}$ erg s$^{-1}$.  Three giant flares with peak luminosities 
$L_{\m{p}}> 10^{44}$ erg s$^{-1}$ 
and durations of a few minutes  were observed from three different SGRs 
(Mazets et al. 1979; Mazets et al. 1999; Hurley et al. 1999; Palmer et al. 2005).  

Energetics of the bursts strongly indicate a magnetar mechanism for these bursts.  Magnetar models (Thompson $\&$ Duncan 1995; 
Thompson $\&$ Duncan 1996) adopt strong  magnetic fields with magnitude 
$B_* > 10^{14}$ G on the stellar surface to explain the burst energetics.
These bursts are likely to be local events near the neutron star and their energies could be originating either from the dipole component or from the higher multipoles of the stellar magnetic field. 
In the current magnetar models, it is the dipole component of the magnetic field which must be of magnetar strength to account for the spin down properties of the AXPs and SGRs. 

In the alternative fallback disk model (Chatterjee, Hernquist, 
$\&$ Narayan 2000; Alpar 2001), the source of the X-rays is accretion onto the neutron star, while the optical/IR light originates from the accretion disk.   Rotational evolution of the neutron star is determined by the interaction between the disk and the dipole component of the neutron star's magnetic field 
($B_*\sim 10^{12}- 10^{13}$ G).  Fallback disk models can account for the period clustering of AXPs and SGRs as the natural outcome of disk-magnetosphere interaction during their lifetimes (Alpar 2001; Ek\c{s}i $\&$ Alpar 2003).  As higher multipole fields rapidly decrease with increasing radial distance (as $r^{-5}$ for quadrupole component), the dipole component of the magnetic field determines the interaction and the angular momentum transfer between the disk and the neutron star.    
The strength of magnetic dipole field in order to explain the period clustering of the AXPs and SGRs over their $\Mdot$ history is 
$B_* \sim 10^{12} - 10^{13}$ G (Alpar 2001; Ek\c{s}i $\&$ Alpar 2003). Therefore, disk models are consistent with the magnetar fields in the quadrupole or higher components, while they are not compatible with magnetar fields in the dipole component. 

Fallback disk models can also explain the enhancements observed in the persistent luminosities of SGRs and AXPs.  
The X-ray enhancement of the SGR 1900+14 following its giant flare can be explained by the relaxation of a disk which has been pushed back by a preceding burst (Ertan $\&$ Alpar 2003). The same model with similar disk parameters can also reproduce the correlated X-ray and IR enhancement of AXP 2259+58, which lasted for $\sim$ 1.5 years, if this is triggered by a burst, with a burst energy estimated to have remained under the detection limits (Ertan, G\"{o}\u{g}\"{u}\c{s} $\&$ Alpar 2006). The suggestion of fallback disks has motivated observational searches for disk emission in the optical and IR bands, and resulted in various constraints on the models.   
Some of the AXPs were observed in more than one IR band
(Hulleman et al. 2001; Israel et al. 2002; Wang $\&$
Chakrabarty 2002; Kaspi et al. 2003; Israel et al. 2003; Hulleman et al. 2004;
Israel et al. 2004; Tam et al. 2004; Morii et al. 2005; Durant $\&$ van Kerkwijk 2006a). 
AXP 4U 0142+61 is the source with most extended observations, as it was 
also observed in the optical R and V bands 
(Hulleman et al. 2000; Hulleman et al. 2004; Dhillon et al. 2005, Wang et al. 2006). 
Discovery of modulation in the R band luminosity of 4U 0142+61 at the neutron star's rotation period P=8.7 s, with a pulsed fraction   
27 \% (Kern $\&$ Mertin 2002; Dhillon et al. 2005), is particularly significant.  This fraction is much higher than the pulsed fraction of 
the X-ray luminosity of this source, indicating that the origin of the pulsed optical emission cannot be the reprocessed X-rays by the disk. 
Magnetospheric models for these pulsations can be built with either a 
dipole magnetar field or within disk-star dynamo model (Cheng $\&$ Ruderman
1991), in which a magnetospheric pulsar activity is sustained by a stellar dipole field of $\sim 10^{12}$ G and a disk protruding within the magnetosphere.  Ertan $\&$ Cheng (2004) showed that this pulsed optical component of the AXP 4U 0142+61 can be explained by both types of magnetospheric models. Thus the presence of strong optical pulsations from the magnetosphere does not rule out the possibility of a fallback disk together with $10^{12}-10^{13}$ G surface dipole magnetic field.   

In the present work, we concentrate on the unpulsed optical/IR emission from the AXPs and SGRs in their persistent states, and test the expectations of the irradiated accretion disk model through the observations in different optical/IR energy bands (V, R, I, J, H, K, and K$_s$).
 The optical/IR emission expected from the irradiated fallback disks was first computed and discussed by Perna, Hernquist \& Narayan  (2000) and Hulleman, van Kerkwijk \& Kulkarni (2000). Using similar irradiation strengths, Perna et al. (2000) and Hulleman et al. (2000) found similar optical fluxes that remain well beyond those indicated by the observations of AXP 4U 0142+61 and AXP 1E 2259+586. To explain this result, Perna et al. (2000) suggested that the inner disk regions could be cut by an advection dominated flow, while Hulleman et al. (2000) concluded that the then existing optical data of the AXP 4U 0142+61 (in I,R,V bands) can only be accounted for by an extremely small outer disk radius, around a few $\times 10^9$ cm (see Ertan et al. 2007 for detailed discussion). In present work, we show that the optical/IR data of the AXPs can be explained by the irradiated accretion disk model without any implausible constraints on the outer and inner disk radii. The main reason for the difference between our results and those of earlier works is that both Hulleman et al. (2000) and Perna et al. (2000) assumed a particular irradiation strength, while we keep it as a free parameter considering the observations of the low mass X-ray binaries (LMXBs) which indicate varying irradiation strengths (see Ertan $\&$ \c{C}al{\i}\c{s}kan (2006) for a detailed discussion)  We give the details of the disk model in \S\ 2. We discuss our results in \S\ 3, and summarize the conclusions in \S\ 4.

\section{Optical/IR Emission from the Irradiated Disk}

Model fits to the contemporaneous X-ray and IR data of the AXP 1E 2259+586 favor the irradiated disk models, though they do not exclude the nonirradiated thin disk model (Ertan, G\"{o}\u{g}\"{u}\c{s} $\&$ Alpar 2006). We start by assuming that the AXP disks are irradiated and include the irradiation strength as a free parameter through our calculations. 

We calculate the disk blackbody emission taking  both the intrinsic dissipation and the irradiation flux into account. A steady disk model is a good approximation for the present evolution of the AXP and SGR disks in their persistent states. For a steady thin disk, the intrinsic dissipation can be written as      
\be
D =\frac{3}{8 \pi} \frac{G M \Mdot}{R^3} 
\label{1}  
\ee  
(see e.g. Frank et al. 2002) 
where $\Mdot$ is the disk mass flow rate, $M$ is the mass of the neutron star and $R$ is the radial distance from the neutron star. In the absence of irradiation, the effective temperature $\Teff$ of the disk is proportional to 
$R^{-3/4}$ for a given $\Mdot$. For an irradiated disk, the irradiation flux  can be written as      
\be
\Firr=\sigma \Tirr^4 = C  \frac{\Mdot c^2}{4 \pi R^2}
\label{2}   
\ee
with
\be
C= \eta (1-\epsilon) \frac{H}{R} \left(\frac{d\ln H}
{d \ln R}-1\right)
\label{3}
\ee 
(Shakura $\&$ Sunyaev 1973), where $\eta$ is the conversion efficiency of the 
rest mass energy into X-rays, $\epsilon$ is the X-ray albedo 
of the disk face, $c$ is the speed of light and $H$ is the pressure scale height of the disk. 
Since the ratio $H/R$ is roughly constant along the disk, the irradiation strength is expected 
to be constant along the disk. Depending on the geometry and  the temperature of the scattering source, efficiency of the irradiation could vary from source to source.    
Estimates for the parameter $C$ are usually in the range 
$10^{-4} - 10^{-3}$ for the low mass X-ray binaries ( de Jong at al. 1996; Dubus et al. 1999; Ertan $\&$ Alpar 2002). 

For a constant $C$ along the disk, the irradiation temperature 
$\Tirr= (\Firr/\sigma)^{1/4}$ is proportional to $R^{-1/2}$ (Eq. 2). 
For small radii, dissipation is the dominant source of the disk emission.  At a critical radius $R_c$, the irradiation flux becomes equal to the dissipation rate, and beyond $R_c$, the dominant heating mechanism is X-ray irradiation. Equating $\Firr$ to $D$ (Eqs. 1 and 2), the critical radius is found to be     
\be
R_c=\frac{3}{2} \frac{GM_*}{C c^2}\simeq  \left(\frac{10^{-4}}{C}\right) 
3\times 10^9 \m{cm}.
\label{4}
\ee  
The effective temperature profile of the disk can be written as
\be 
\Teff = \L(\frac{D}{\sigma}+\frac{\Firr}{\sigma}\R)^{1/4}
\label{5}
\ee
where $\sigma$ is the Stefan-Boltzmann constant. 

To find the model disk flux in a given observational band, we integrate the calculated blackbody emissions of all radial grids radiating in this band. For comparison with data, we calculate the model disk fluxes along the optical/IR bands V, R, H, I, J, K and K$_s$.  For all sources we set $\cos i=1$ where $i$ is the angle between the disk normal and the line of sight of the observer. We equate the disk mass flow rate $\Mdot$ to the accretion rate onto the neutron star, thus assuming the mass loss due to the propeller effect is negligible. We first adjust $\Mdot$ to obtain the observed X-ray flux. Next, using this value of $\Mdot$ and taking the strength of the magnetic dipole field $B_*= 10^{12}$ G on the surface of the neutron star we calculate the Alfv$\acute{e}$n radius $R_\m{A}$ which we take to be the inner radius of the disk. Then, we look for a good fit to the overall available optical/IR data by adjusting the irradiation strength $C$  within the uncertainties discussed in \S\ 3. For the sources with more than one detections in a particular IR/optical band we take the observation nearest to the date of X-ray observation.

\section{Results and Discussion} 

We summarize our results in Table 1.  For each source, the first column  gives the unabsorbed flux data obtained from the observed magnitudes and the estimated  $\Av$ values (for references see Table 1), and the second column gives the model fluxes. For the AXP 4U 0142+61, reasonable range of reddening is $2.6 < \Av < 5.1$ (Hulleman et al. 2004). We obtain a good fit with $\Av=3.5$ (Fig. 1). Table 1 shows that the irradiated steady disk model is also in agreement with all the other AXPs observed in the optical and IR bands. The parameters of the model for each source are given in Table 2.  

At present, AXP 4U 0142+61, which has been observed in nine different optical/IR bands from  8 $\mu$m to V in the same X-ray luminosity regime, seems to be the best source to study the properties of AXPs in the persistent state.    
The irradiation parameter $C$ obtained from our model fits remains in the range
($10^{-4} < C < 10^{-3}$) estimated from the observations of LMXBs and the disk stability analyses of the soft X-ray transients (de Jong et al. 1996;  Dubus et al. 1999; Ertan $\&$ Alpar 2002). Within the critical radius $R_\m{c}$ given by Eq. 4, dissipation is the dominant heating mechanism. For the disk model of the AXP 4U 0142+61, $R_\m{c}\simeq 3\times 10^9$ cm and $\Rin = 1\times 10^9$ cm. The innermost disk emitting mostly in the UV bands also contributes to the optical emission. The radial distance at which the disk blackbody temperatures peak at the optical bands (R,V) is about $10^{10}$ cm. Peak temperatures of the IR bands from I to SPITZER mid IR bands (4.5 $\mu$m and 8 $\mu$m) lie between $R\sim 2\times 10^{10}$ cm and $R\sim 10^{12}$ cm. There are several observations of this source in some of the IR bands; we adopt the data nearest to the X-ray observation epoch used in our analysis here. 

For AXP J1708-40, Durant and van Kerkwijk (2006a) recently found that the previously reported IR data in $\Ks$, H, and J bands are likely to be a background star. They found another object within the positional error cycle and argued that this second object is more likely to be the IR counterpart to the AXP J1708-40. For this source, we adopt the IR ($\Ks$, H, J) data set reported by  Durant and van Kerkwijk (2006a).

For the the AXP 1E 2259+586, we use the data taken before the X-ray enhancement phase of this source (Hulleman et al. 2001). This source was detected in  $\Ks$ band and there are upper limits for I and R bands. Our model flux values are three and ten times below the upper limits reported for I and R bands respectively. 

AXP 1E 1048-59 was detected in  $\Ks$, H and I bands (Wang $\&$ Chakrabarty 2002). Observed X-ray flux from this source between December 2000 to January 2003 show a variation within a factor of 5 (Mereghetti et al. 2004). We use the X-ray flux obtained from  the nearest X-ray observation to the date of the IR observations when the source was in the persistent state. 

AXP 1E 1841 was detected only in the $\Ks$ band, and there is a high upper limit in the R band (Wachter et al. 2004). Model estimates in other optical/IR bands for this source (and the other AXPs) can be tested by future optical and IR  observations.

We note that our model does not address the uncertainties in the innermost disk interacting with the magnetosphere, the contributions from the  magnetospheric pulsed emission which is known to have a fraction about 27 \% in the R band, and the shielding effects which might decrease the irradiation strength at some regions of the disk. All of our optical and IR flux values remain within 30 \% of the data points, which is a reasonable fit considering the uncertainties mentioned above.

\clearpage
\begin{figure*}
\vspace{-8 cm}
\includegraphics[width=1.00\textwidth]{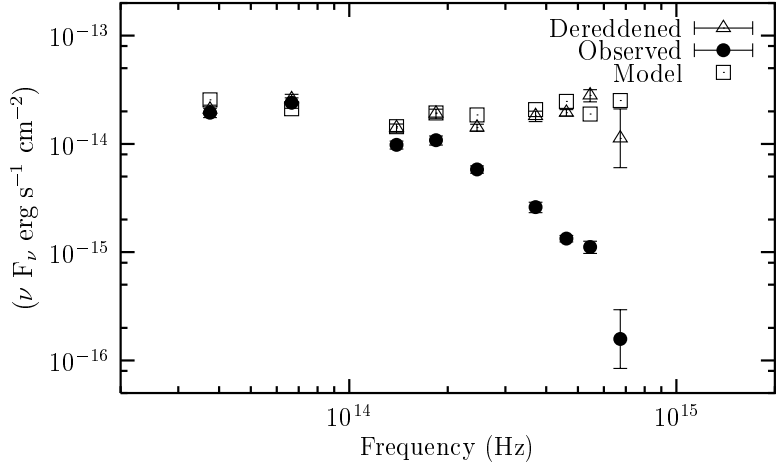}
\vspace{-10 cm}
\caption{Energy flux data and irradiated disk model values for
the  AXP 4U 0142+61 in the optical and infrared bands (B, V, R, I, J, H, K$_S$,
4.5 $\mu$m and 8 $\mu$m). Filled circles are the observed (absorbed) data (taken from Hulleman et al. 2000  (V, R, I), Hulleman et al. 2004 (B,K$_S$), Morii et al. 2005 (H, J),  Wang et al. 2006 (4.5 $\mu$m and 8 $\mu$m), and triangles are dereddened data with $A_V=3.5$. Squares are the irradiated disk model energy flux values (see \S\ 2 for details).}   
\label{fig:1}
\end{figure*}

\begin{table*}
\vspace{1.50cm} 
\caption{The data flux values were calculated by using the magnitudes and $\Av$ values given in the references below. REFERENCES: (J1708-40) Durant \& van Kerkwijk 2006a, Rea et al. 2003; (1E 2259+586) Hulleman et al. 2001, Woods et al.  2004; (1E 1841-45) Wachter et al. 2004, Morii et al. 2003; (1E 1048-59) Wang $\&$ Chakrabarty 2002, Mereghetti et al. 2004 }

\begin{tabular}{l|cc|cc|cc|cc}
\hline
&\multicolumn{2}{c|}{J1708-40}&\multicolumn{2}{c|}{1E 2259+58} 
&\multicolumn{2}{c|}{1E 1841-045}
&\multicolumn{2}{c}{1E 1048-59}
\\
\hline
&\multicolumn{2}{c|}
{Flux}&\multicolumn{2}{c|}
{Flux }&\multicolumn{2}{c|}
{Flux }&\multicolumn{2}{c}
{Flux }
\\
&\multicolumn{2}{c|}
{($10^{-15}$erg s$^{-1}$ cm$^{-2}$) }&\multicolumn{2}{c|}
{($10^{-15}$erg s$^{-1}$ cm$^{-2}$)}&\multicolumn{2}{c|}
{($10^{-15}$erg s$^{-1}$ cm$^{-2}$) }&\multicolumn{2}{c}
{($10^{-15}$erg s$^{-1}$ cm$^{-2}$) }
\\
\hline
Band&Data&Model &Data&Model &Data&Model&Data&Model 
\\
&$(\Av=7.8)$&&$(\Av=6.1)$&&$(\Av=8.4)$&&$(\Av=5.6)$&
\\
\hline
K$_s$&49&44&3.7 &3.6  &68 &68 &29&22  
\\
\hline
K  &  & 54&     &4.5  &  & 84 & &27  
\\
\hline
H&51 &57 &     &4.8  &  &89 &22&28  
\\
\hline
J&50 &53 &     &4.4  &  &83 &33&26  
\\
\hline
I&   &56&$<$15  &4.4 &  &88 & & 24 
\\
\hline
R&   &65 &$<$42 &4.5 &$<3.8\times10^5$&100 & &26  
\\
\hline
V&   &48 &      &3.0 &  &76 & &18  
\\
\hline

\end{tabular}

\end{table*}

\begin{table*}
\vspace{1.5cm} 
\begin{center}
\caption{The parameters of the irradiated disk model which gives the optical/IR flux values seen in Table 1. For all the sources, we set $\cos i=1$ where $i$ is the inclination angle between the disk normal and the line of sight of the observer, and we take the outer disk radius $\Rout=2\times 10^{12}$ cm. See Section 3 for details}

\begin{tabular}{l|c||c||c||c|c}
\hline
&1RXS J1708-40&1E 2259+58 &4U 0142+61&1E 1841-045
&1E 1048-59
\\
\hline
$\Rin$ (cm)&$1.2\times 10^9 $& $2.3\times 10^9$
&$1.0\times 10^9$&$1.3\times 10^9$&
$3.3\times 10^9$   
\\
\hline
$C$&$5.0\times 10^{-4}$&$ 1.6\times 10^{-4}$
&$1.0\times 10^{-4}$&$7.2\times 10^{-4}$&
$7.0\times 10^{-4}$    
\\
\hline
$d$(kpc)& 5& 3& 3&7 &3   
\\
\hline
$\Mdot$ (g s$^{-1})$&$1.0\times 10^{15}$&$9.1\times 10^{13}$&$4.8\times 10^{14}$&$2.2\times 10^{15}$&$ 1.3\times 10^{14}$
\\
\hline

\end{tabular}
\end{center}
\end{table*}

\clearpage
Since there is no detections in short wavelength optical bands for the AXPs (except for AXP 4U 0142+61), model fits are not sensitive to the chosen inner disk radii. For the model fits, we equate the inner disk radii to the
 Alfv$\acute{\m{e}}$n
radii (Table 2) corresponding to a magnetic field with magnitude $B_* = 10^{12}$ G on the stellar surface and the accretion rates derived from the estimated X-ray luminosities (see Table 1 for references).  
On the other hand, optical data of the AXP 4U 0142+61 in R and V bands provide a constraint for the inner disk radius, and thereby for the strength of the magnetic dipole field of this source (see \S\ 4).

\section{Conclusion}

We have shown that the optical, infrared and X-ray observations of the AXPs in their persistent states can be explained with irradiated disk models.  Among the AXPs, 4U 0142+61 is currently the only source which provides  an upper limit for the inner disk radius through its optical (R,V) data. For the best model fit for this source, which we have obtained with  $\Av =3.5$,  the model inner disk radius ($\sim 10^9$ cm) is around the Alfv$\acute{e}$n radius for the accretion rate, estimated from the X-ray luminosity, together with a dipole magnetic field strength  $B_* \simeq 10^{12}$ G on the neutron star surface. Nevertheless, it is possible to obtain reasonable fits by increasing the inner disk radius and decreasing the reddening accordingly. For $\Av=2.6$, the minimum value of the reddening in the reasonable range 
$2.6 < \Av < 5.1$ (Hulleman et al. 2004), we obtain the best fit with 
$\Rin \simeq 8\times 10^9$ cm which corresponds to the maximum reasonable dipole field strength  
$B_*\simeq 4\times 10^{13}$ G on the pole and half of this on the equator of the neutron star. Nevertleless, we note that these limits could be increased depending on the amount of possible mass loss due to propeller effect and/or on how much the inner disk radius penetrates inside the Alfv$\acute{e}$n radius (see Ertan et al. 2007 for a detailed discussion for 4U 0142+61). On the other  hand, even including these possibilities, very recent analysis concluding $\Av=3.5\pm 0.4$ (Durant \& van Kerkwijk 2006b) implies surface dipole magnetic field strengths less than about $10^{13}$ G.
The magnetar fields ($B_* > 10^{14}$ G) in multipoles, that could be responsible for the burst energies, are compatible with this picture, while the optical data excludes a hybrid model involving a disk surrounding a dipole field of magnetar strength. 

On the other hand, existing IR data of the AXPs, including recent observations of 4U 0142+61 by SPITZER in 4.5 $\mu$m and 8 $\mu$m bands (Wang et al. 2006), do not put an upper limit for the extension of the outer disk radius $\Rout$. The lower limit for $\Rout$ provided by the longest wavelength IR data of the AXP 4U 0142+61 is around $10^{12}$ G.   
Further observations in the longer wavelength infrared bands by SPITZER space telescope will provide valuable information about the structure and possibly the extension of the fallback disks around these systems.



\begin{thebibliography}{99}
%
%


\bibitem{} Alpar, M.A. ApJ, {\bf 554}, 1245 (2001)
\bibitem{} Chatterjee, P., Hernquist, L., $\&$ Narayan, R.  ApJ, {\bf 534}, 373 (2000)
\bibitem{575} Cheng, K.S., $\&$ Ruderman, M. Apj, {\bf 373}, 187 (1991) 
\bibitem{572} de Jong, J. A., van Paradijs, J., $\&$ Augusteijn, T. A$\&$A, {\bf 314}, 484 (1996)
\bibitem{279} Dhillon, V.S., Marsh, T.R., Hulleman, et al.  MNRAS, {\bf 363}, 609 (2005) 
\bibitem{614} Dubus, G., Lasota, J.-P., Hameury, J.-M., $\&$ Charles, P, 
 MNRAS, {\bf 303}, 139 (1999)
\bibitem{Durant06} Durant, M. \& van Kerkwijk, M. H.  ApJ, {\bf 648}, 534 (2006a)
\bibitem{Durant06} Durant, M. \& van Kerkwijk, M. H. accepted for publication in ApJ (astro-ph/0606604) (2006b)
\bibitem{620} Ek\c{s}i, Y. K., $\&$ Alpar, M.A. ApJ, {\bf 559}, 450 (2003)
\bibitem{288} Ertan, {\"U}., $\&$  Alpar, M.A. A$\&$A, {\bf 393}, 205 (2002)
\bibitem{74} Ertan, {\"U}., $\&$  Alpar, M.A.  ApJ, {\bf 593}, L93 (2003)
\bibitem{54} Ertan, {\"U}., $\&$ Cheng, K.S. ApJ, {\bf 605}, 840 (2004)   
\bibitem{57} Ertan, {\"U}., $\&$  \c{C}al{\i}\c{s}kan, \c{S}. ApJ, {\bf 649}, L87 (2006)
\bibitem{574} Ertan, {\"U}, Erkut, M. H., Ek\c{s}i, Y. K., $\&$ Alpar, M.A.  (accepted for publication in Apj, astro-ph/0606259)(2007)
\bibitem{EGA06} Ertan, \"{U}., G\"{o}\u{g}\"{u}\c{s}, E., \& Alpar, M.A. ApJ, {\bf 640}, 435 (2006)  
\bibitem{575} Feroci, M., Hurley, K., Duncan, R., $\&$ Thompson, C. ApJ, 
{\bf 549}, 1021 (2001)
\bibitem{577} Frank, J., King, A.R., $\&$ Raine, D.  Accretion 
Power in Astrophysics (Cambridge: Cambridge University Press) (2002)
\bibitem{630}Hulleman, F., van Kerkwijk, M. H., $\&$ Kulkarni, S.R. Nature, {\bf 408}, 689 (2000)
\bibitem{632} Hulleman, F., Tennant,A.F.,van Kerkwijk, M.H.,
  et al., Apj, {\bf 563}, L49 (2001)
\bibitem{634} Hulleman, F., van Kerkwijk, M.H., Kulkarni, S.R. A$\&$A, {\bf 416}, 1037  (2004)
\bibitem{636}Hurley, K., Cline, T., Mazets, E., et al. Nature, {\bf 397}, 41 (1999)
\bibitem{641} Hurley, K. in AIP Conf. Proc. {\bf 526}, Gamma-Ray Bursts:
   Fifth Huntsville Symp., ed. R. M. Kippen, R. S. Mallozzi, $\&$
   G. J. Fishman (New York: AIP), 763  (2000)
\bibitem{644} Israel, G. L.,Covino, S., Perna, R., et al.  ApJ, 
  {\bf 589}, L93 (2003)
\bibitem{648} Israel, G. L.,Covino,S., Stella, L., et al.  ApJ, {\bf 580}, L143 (2002)
\bibitem{652} Israel, G. L., Rea, N., Mangano, V., et al. ApJ, {\bf 603}, L97 (2004)
\bibitem{657} Kaspi, V.M. et al. ApJ, {\bf 588}, L93 (2003)
\bibitem{658} Kern B. $\&$ Martin, C. Nature, {\bf 417}, 527
\bibitem{582} Mazets, E.P., Cline, T., Aptekar, R.L., et al.  Astron.Lett., {\bf 25}, 635 (1999)
\bibitem{585} Mazets, E.P., Golenetskii, S.V., Il'inskii, V.N.,
   Aptekar, R.L., $\&$ Guryan, Y. A.  Nature, {\bf 282}, 587   
\bibitem{} Mereghetti, S., Tiengo, A., Stella, L. et al. ApJ, {\bf 608}, 427 (2004) 
\bibitem{669} Mereghetti, S., Chiarlone, L., Israel, G. L., $\&$ Stella, 
L. in Proc. 270th WE-Heraus Seminar on Neutron Stars, Pulsars and 
Supernova Remnants, ed. W. Becker, H. Lecsch, $\&$ J. Trumper (MPE Rep. 278; Garching: MPE), 29 (2002)
\bibitem{} Morii, M., Sato, R., Kataoka, J., $\&$ Kawai, N. PASJ, 
{\bf 55}, L45 (2003) 
\bibitem{Morii05} Morii, M., Kawai, N., Kataoka, J., et al. Advances in Space Research, {\bf 35}, 1177  (2005)  
\bibitem{334}Palmer, D.M. et al, Nature, {\bf 434}, 1107 (2005)
\bibitem{} Patel, S.K., et al. ApJ, {\bf 587}, 367 (2003)
\bibitem{PHN} Perna, R., Hernquist, L., \& Narayan, R., Apj, {\bf 541}, 344 (2000)
\bibitem{} Rea, N., et al. ApJ, {\bf 586}, L65 (2003)
\bibitem{590} Shakura, N.I., $\&$ Sunyaev, R.A., A$\&$A, {\bf 24}, 337 (1973)
\bibitem{590} Tam, C. R., Kaspi, V.M., van Kerkwijk, M.H., $\&$ 
    Durant, M. ApJ, {\bf 617}, L53 (2004)
\bibitem{591} Thompson, C., $\&$ Duncan, R. C. MNRAS, {\bf 275}, 255 (1995)
\bibitem{591} Thompson, C., $\&$ Duncan, R. C. ApJ, {\bf 473}, 322 (1996)
\bibitem{} Wachter, S., Patel, S.K., Kouveliotou, C. et al. ApJ, {\bf 615}, 887 (2004)
\bibitem{WCK06} Wang, Z., Chakrabarty, D.  \& Kaplan, D. Nature, 
{\bf 440}, 772 (2006)
\bibitem{591} Wang, Z., $\&$ Chakrabarty, D. ApJ, {\bf 579}, L33 (2002)
\bibitem{595} Woods, P. M. et al. ApJ, {\bf 605}, 378 (2004) 
\bibitem{595} Woods, P. M. $\&$ Thompson, C. in "Compact Stellar X-ray Sources", eds, W.H.G.Lewin $\&$ M. van der Klis  (2006)  
\end{thebibliography}


\end{document}